\documentclass[longauth,traditabstract]{aa} 

\usepackage{graphicx}
\usepackage{txfonts}
\usepackage{epsfig}
\begin{document}

\title{AGILE detection of extreme $\gamma$-ray activity from the blazar PKS
  1510-089 during March 2009. Multifrequency analysis\thanks{The
  radio-to-optical data presented in this paper are stored in the GASP-WEBT
  archive; for questions regarding their availability, please contact the WEBT President Massimo Villata.}}

\author{F.~D'Ammando\inst{1,2,3}, C.~M.~Raiteri\inst{4}, M.~Villata\inst{4},
  P. Romano\inst{1}, G.~Pucella\inst{5}, H.~A.~Krimm\inst{6},
  S. Covino\inst{7,8}, M.~Orienti\inst{9}, G.~Giovannini\inst{9},
  S.~Vercellone\inst{1}
% ...
}

\institute{$^{1}$ INAF, Istituto di Astrofisica Spaziale e Fisica Cosmica, Via U.~La Malfa 153, I-90146 Palermo, Italy \\
           $^{2}$ Dip. di Fisica, Univ. di Roma ``Tor Vergata'', Via della Ricerca Scientifica 1, I-00133 Roma, Italy \\  
           $^{3}$ INAF/IASF--Roma, Via Fosso del Cavaliere 100, I-00133 Roma, Italy \\ 
           $^{4}$ INAF, Oss. Astronomico di Torino, Via Osservatorio 20,
           I-10025 Pino Torinese (Torino), Italy \\
           $^{5}$ ENEA--Frascati, Via E. Fermi 45, I-00044 Frascati (Roma),
           Italy \\ 
           $^{6}$ NASA / Goddard Space Flight Center, Greenbelt, MD 20771, USA \\
           $^{7}$ INAF, Oss. Astronomico Brera, Via Bianchi 46, I-23807 Merate
           (LC), Italy \\
           $^{8}$ INAF/TNG Fundaci\'on Galileo Galilei, Rambia Jos\'e Ana
           Fern\'andez P\'erez 7, 38712 Brena Baja, Tenerife Spain \\
           $^{9}$ INAF-Istituto di Radioastronomia, Via P. Gobetti 101,
           I-40129, Bologna, Italy\\
           $^{10}$ Scuola Normale Superiore, Piazza dei Cavalieri 7, I-56126 Pisa, Italy \\ 
           $^{11}$ INAF Osservatorio Astronomico di Trieste, Via G. Tiepolo 11, I-34143 \\
           $^{12}$ Dip. di Fisica and INFN Trieste, Via Valerio 2, I-34127 Trieste, Italy \\ 
           $^{13}$ INFN--Pavia, Via Bassi 6, I-27100 Pavia, Italy \\ 
           $^{14}$ Dip. di Fisica Nucleare e Teorica, Univ. di Pavia, Via Bassi 6, I-27100 Pavia, Italy \\
           $^{15}$ INAF/IASF--Bologna, Via Gobetti 101, I-40129 Bologna, Italy \\ 
           $^{16}$ INAF/IASF--Milano, Via E.~Bassini 15, I-20133 Milano, Italy
           \\
           $^{17}$ Dip. di Fisica Generale dell'Universit\'a, Via P.~Giuria 1,
           I-10125 Torino, Italy \\
           $^{18}$ ENEA--Bologna, Via dei Martiri di Monte Sole 4, I-40129 Bologna, Italy \\ 
           $^{19}$ INFN--Roma ``La Sapienza'', Piazzale A. Moro 2, I-00185 Roma, Italy \\
           $^{20}$ INFN--Roma ``Tor Vergata'', Via della Ricerca Scientifica 1, I-00133 Roma, Italy \\
           $^{21}$ INAF--Oss. Astronomico di Cagliari, loc. Poggio dei Pini, strada 54, I-09012 Capoterra (CA), Italy \\ 
           $^{22}$ Dip. di Fisica, Univ. dell'Insubria, Via Valleggio 11, I-22100 Como, Italy \\ 
           $^{23}$ Institute for Astrophysical Research, Boston University,
           725 Commonwealth Avenue, Boston, MA 02215, USA \\
           $^{24}$ Instituto de Astrof\'{\i}sica de Andaluc\'{\i}a, CSIC,
           Apartado 3004, 18080 Granada, Spain \\
           $^{25}$ Department of Astronomy, University of Michigan, Ann Arbor,
           MI 48109, USA\\
           $^{26}$ Pulkovo Observatory, Russian Academy of Sciences, 196140, St.-Petersburg, Russia \\         
           $^{27}$ MPIfR, D-53121 Bonn, Germany \\
           $^{28}$ Instituto de Astronomia, Universidad Nacional Autonoma de Mexico, Mexico, D. F. Mexico \\
           $^{29}$ Tuorla Observatory, Department of Physics and Astronomy, University of Turku, V\"ais\"al\"antie 20, 21500 Piikki\"o, Finland \\
           $^{30}$ Astron.\ Inst., St.-Petersburg State Univ., 198504 St.-Petersburg, Russia \\
           $^{31}$ INAF--Osservatorio Astrofisico di Catania, Via S. Sofia 78, I-95123 Catania, Italy\\ 
           $^{32}$ Institute of Astronomy, National Central University, Taiwan\\
           $^{33}$ Lulin Observatory, Institute of Astronomy, National Central
           University, Taiwan \\ 
           $^{34}$ INAF, Osservatorio Astronomico di Roma, Via di Frascati 33,
           I-00040, Monte Porzio Catone, Italy \\ 
           $^{35}$ INAF, Osservatorio Astronomico di Collurania, Via Mentore Maggini, I-64100 Teramo, Italy\\
           $^{36}$ Agrupaci\'o Astron\`omica de Sabadell, Spain \\
           $^{37}$ Harvard-Smithsonian Center for Astrophysics, Cambridge, Garden st. 60, MA 02138, USA \\
           $^{38}$ School Of Cosmic Physics, Dublin Institute For Advanced
           Studies, Ireland \\
           $^{39}$ ZAH, Landessternwarte Heidelberg, K\"onigstuhl, D-69117 Heidelberg, Germany \\
           $^{40}$ Instituto de Astronomia, Universidad Nacional Autonoma de
           Mexico, 2280 Ensenada, B.C. Mexico \\
           $^{41}$ Aalto University Mets\"ahovi Radio Observatory,
           Mets\"ahovintie 114, FIN-02540 Kym\"al\"a, Finland \\
           $^{42}$ Department of Physics, Purdue University, 525 Northwestern
           Ave., West Lafayette, IN 47907, USA \\          
           $^{43}$ Abastumani Observatory, 383762 Abastumani, Georgia \\
           $^{44}$ Isaac Newton Institute of Chile, St.-Petersburg Branch \\
           $^{45}$ School Of Cosmic Physics, Dublin Institute For Advanced
           Studies, Ireland \\
           $^{46}$ Department of Physics and Astronomy, University of Southampton S017 1BJ, UK \\
           $^{47}$ Finnish Centre for Astronomy with ESO (FINCA), University of Turku, FI-21550 Piikki\"o, Finland \\
           $^{48}$ ASI--ASDC, Via G. Galilei, I-00044 Frascati (Roma), Italy \\
           $^{49}$ ASI, Viale Liegi 26, I-00198 Roma, Italy \\}
\offprints{F.~D'Ammando, \email{dammando@ifc.inaf.it} }

  \date{received; accepted}


  \abstract{We report on the amazing $\gamma$-ray activity from the flat spectrum radio quasar (FSRQ) PKS 1510$-$089 observed by the AGILE satellite in
  March 2009. In the same period a radio-to-optical monitoring of the source
  was provided by the GASP--WEBT and REM facilities. In the radio band we made
  use also of multi-epoch 15-GHz Very Long Baseline Array data from the MOJAVE
  Program to get information on the parsec-scale structure. Moreover, several
  {{\it Swift}} target of opportunity observations were triggered, adding
  important information on the source behaviour from optical/UV to hard
  X-rays. We paid particular attention to the calibration of the {\it
  Swift}/UVOT data to make it suitable to the blazars spectra. Simultaneous
  observations from radio to $\gamma$-rays allowed us to study in detail the
  correlation among the emission variability at different frequencies and to
  investigate the mechanisms at work during this high activity state of the source.

 In the period 9--30 March 2009, AGILE detected $\gamma$-ray emission from PKS
   1510$-$089 at a significance level of 21.5-$\sigma$ with an average flux over the entire period of $(311 \pm 21) \times 10^{-8}$ photons
   cm$^{-2}$ s$^{-1}$ for photon energies above 100 MeV, and a peak level of $(702 \pm 131) \times 10^{-8}$ photons cm$^{-2}$ s$^{-1}$ on daily
   integration. The activity detected in $\gamma$-rays occurred during a
   period of increasing activity from near-infrared to UV, as monitored by
   GASP--WEBT, REM and {{\it Swift}}/UVOT. A flaring episode on 26-27 March
   2009 was detected from near-IR to UV, suggesting that a single mechanism is
   responsible for the flux enhancement observed at the end of March. By
   contrast, {\it Swift}/XRT observations seem to show no clear correlation of
   the X-ray fluxes with the optical and $\gamma$-ray ones. However, the X-ray
   observations show a harder photon index ($\Gamma_{\rm x}$$\simeq$ 1.3--1.6)
   with respect to most FSRQs and a hint of harder-when-brighter behaviour,
   indicating the possible presence of a second emission component at soft
   X-ray energies. Moreover, the broad band spectrum from radio-to-UV
   confirmed the evidence of thermal features in the optical/UV spectrum of
   PKS 1510$-$089 also during high $\gamma$-ray state. On the other hand,
   during 25-26 March 2009 a flat spectrum in the optical/UV energy band was
   observed, suggesting an important contribution of the synchrotron emission
   in this part of the spectrum during the brightest $\gamma$-ray flare,
   therefore a significant shift of the synchrotron peak.}  
  
\authorrunning{F.~D'Ammando et al.}
\titlerunning{AGILE detection of extreme $\gamma$-ray activity from the blazar
  PKS 1510$-$089 during March 2009}
\maketitle

\section{Introduction}

According to the current paradigm (see Urry $\&$ Padovani 1995) Active
Galactic Nuclei (AGNs) are galaxies whose emission is dominated by a bright
central core, including a super massive black hole as central engine,
surrounded by an accretion disk and by fast-moving clouds under the influence
of the strong gravitational field, emitting Doppler broadened lines. Absorbing
material in a flattened configuration, idealized as a dust torus, obscures the
central parts so that for transverse line of sight only the more distant
narrow-line emitting clouds are seen directly. In radio-loud objects we have
the additional presence of a relativistic jet, roughly perpendicular to the disk. Within this scheme, blazars represent the fraction of AGNs with their
jet at smaller angles with respect to our line of sight, which causes relativistic aberration and emission amplification (Blandford $\&$ Rees 1978).

Blazars are the most enigmatic subclass of AGNs, characterized by the emission of strong non-thermal radiation across the entire electromagnetic
spectrum, from radio to very high $\gamma$-ray energies. Because of their
peculiar properties and amplified emission, blazars offer the unique
possibility to probe the central region of AGNs, and shed light on the
mechanism responsible for the extraction of energy from the central black hole
and for the acceleration and collimation of relativistic electrons into
jets. With the detection of several blazars in the $\gamma$-rays by EGRET
(Hartman et al.~1999) the study of this class of objects has made significant
progress but, despite the big efforts devoted to the investigation of the
mechanisms responsible for the emission in blazars, the definitive answer is
still missing. The new generation of high-energy space missions, the {\it
  Astrorivelatore Gamma a Immagini LEggero} (AGILE) and {\it Fermi} Gamma-ray Space Telescope (GST) satellites, have to address some of the fundamental issues that were left unresolved by EGRET.

PKS 1510$-$089 is a nearby (z=0.361) blazar belonging to the class of the flat
spectrum radio quasars (FSRQs) with radiative output dominated by the
$\gamma$-ray emission, while the synchrotron emission peaks around IR
frequencies, below a pronunced UV bump, likely due to the thermal emission
from the accretion disk (Malkan $\&$ Moore 1986; Pian $\&$ Treves 1993; D'Ammando et al. 2009a). 

Its radio emission exhibits very rapid, large amplitude variations in both
total and polarized flux (Aller, Aller, $\&$ Hughes 1996). Moreover, the radio
jet shows superluminal motion with an apparent velocity higher than 20$c$, among the fastest of all blazars observed thus far (Jorstad et al.~2005), with
the parsec and kiloparsec scale jets apparently misaligned by $\sim$180 degrees (Homan et al.~2002, Wardle et al.~2005).

PKS 1510$-$089 has been extensively observed in the X-rays during the last
three decades, since the observations by the {\it Einstein} satellite in 1980s
(Canizares $\&$ White 1989). The X-ray spectrum as observed by ASCA in the
2--10 keV band was very hard with photon index of $\Gamma$ $\simeq$ 1.3
(Singh, Shrader $\&$ George 1997), but steepened ($\Gamma$ $\simeq$ 1.9) in
the ROSAT bandpass (0.1--2.4 keV), suggesting the presence of a spectral break
around 1--2 keV, associated with the existence of a soft X-ray excess (Siebert
et al.~1998). Subsequent observations by {\it Beppo}SAX (Tavecchio et
al.~2000) and {\it Chandra} (Gambill et al.~2003) confirmed the presence of a
soft X-ray excess below 1 keV. Evidence of a similar soft X-ray excess has
been detected in other blazars such as 3C 273, 3C 279, AO 0235+164, and 3C
454.3. The origin of this excess is still an open issue, not only for blazars
but for all AGNs (see e.g.~D'Ammando et al.~2008, and the references therein).

The interest in the peculiar X-ray spectrum of the source led up to a
monitoring campaign organized during August 2006 by {\it Suzaku} and {\it
  Swift}. The {\it Suzaku} observations confirmed the presence of a soft X-ray
excess, suggesting that it could be a feature of the bulk Comptonization,
whereas the {\it Swift}/XRT observations revealed significant spectral
evolution of the X-ray emission on timescales of a week: the X-ray spectrum
becomes harder as the source gets brighter (Kataoka et
al.~2008). Unfortunately no $\gamma$-ray satellites were operating at that
time, therefore it was not possible to investigate the correlation between
X-ray and $\gamma$-ray behaviour. However, $\gamma$-ray emission from PKS
1510$-$089 had already been detected in the past by EGRET during
low/intermediate states and was found to be only slightly variable, with an
integrated flux above 100 MeV varying between (13 $\pm$ 5) and (49 $\pm$ 18) $\times$ 10$^{-8}$ photons cm$^{-2}$ s$^{-1}$ (Hartman et al.~1999). 

Instead, in the last three years PKS 1510$-$089 showed extreme variability
over the whole electromagnetic spectrum and in particular very intense
$\gamma$-ray activity was detected, intermittently, by the AGILE and {\it
  Fermi} satellites. The monitoring of the source between September 2008 and June 2009 with the Large Area Telescope (LAT) onboard {\it Fermi}
is summarized in Abdo et al.~(2010a). The Gamma-Ray Imaging Detector (GRID) onboard AGILE detected flaring episodes in August 2007 (Pucella
et al.~2008) and March 2008 (D'Ammando et al.~2009a). Subsequently, an extraordinary $\gamma$-ray activity was detected in March 2009, with several
flaring episodes and a flux that reached $\sim$ 700 $\times$ 10$^{-8}$ photons
cm$^{-2}$ s$^{-1}$. Preliminary results were presented in D'Ammando et
al.~(2009b), Pucella et al.~(2009), Vercellone et al.~(2009). Recently,
Marscher et al.~(2010) studying the flux behaviour at different frequencies,
linear optical polarization, and parsec-scale structure of PKS 1510$-$089
found a correspondence between the rotation of the optical polarization angle and the $\gamma$-ray activity during the first six months of 2009.

\noindent In this paper we discuss the results of the analysis of the multiwavelength data of PKS 1510$-$089 collected by GLAST-AGILE Support
Program (GASP) of the Whole Earth Blazar Telescope (WEBT), Rapid Eye Mount
(REM), {\it Swift} and AGILE during March 2009 and their implications for the
emission mechanisms at work in this source. The paper is organized as
follows. Section 2 briefly introduces the multiwavelength coverage on PKS
1510$-$089. In Sections 3 through 7 we present the analysis and results of
AGILE, {\it Swift}, GASP-WEBT, REM, and Very Long Baseline Array (VLBA) data,
respectively. In Section 8 we discuss the contribution of the thermal emission
in the low-energy part of the spectrum and the correlation between the
emission at different energy bands. Finally, in Section 9 we draw our conclusions.

Throughout this paper the quoted uncertainties are given at 1-$\sigma$ level,
unless otherwise stated, and the photon indices are parameterized as $N(E)
\propto E^{-\Gamma}$ (ph cm$^{-2}$ s$^{-1}$ keV$^{-1}$ or MeV$^{-1}$ ) with
$\Gamma = \alpha +1$ ($\alpha$ being the spectral index). 

\section{The multiwavelength coverage}

The high $\gamma$-ray activity observed by AGILE during March 2009 triggered
15 {\it Swift} ToO observations, starting from 11 March 2009, for a total of
46 ks. Moreover, the monitoring by GASP--WEBT in February--March 2009 provided
important information from radio-to-optical band and, together with the data
collected in near-IR and optical bands by REM during March 2009, allowed us to
also obtain an excellent coverage in the low-energy part of the broad band
spectrum of the source. Finally, high resolution radio VLBI data at 15 GHz
were obtained by monitoring the source within the Monitoring of Jets in Active galactic nuclei with VLBA Experiment (MOJAVE) project. 

A summary of the complete multiwavelength data set on PKS 1510$-$089 presented in this paper can be found in Table 1. 

\begin{table}[!hhh]
\begin{center}
\begin{tabular}{ccc}
\hline \hline
\label{table:Observatories}
Waveband & Observatory  & Frequency/Band  \\ 
\hline
Radio & SMA & 230 GHz \\
& Noto & 43 GHz \\
& Mets\"ahovi & 37 GHz \\
& VLBA & 15, 43 GHz \\
& UMRAO & 8.0, 14.5 GHz  \\
& Medicina & 5, 22 GHz \\
\hline
Near-IR & REM & $J$, $H$, $K$ \\
& Campo Imperatore & $J$, $H$, $K$ \\
& Roque (Liverpool) & $H$ \\
\hline
Optical & Abastumani & $R$ \\
& Calar Alto & $R$ \\
& Castelgrande & $B$, $V$, $R$, $I$ \\
& L'Ampolla & $R$ \\
& La Silla (MPG/ESO) & $B$, $V$, $R$, $I$ \\
& Lulin (SLT) & $R$ \\
& Roque (KVA and Liverpool) & $R$ \\
& San Pedro Martir & $R$ \\
& St. Petersburg & $B$, $V$, $R$, $I$ \\ 
& Valle d'Aosta & $R$ \\
& REM & $V$, $R$, $I$ \\
& {\it Swift}/UVOT & $v$, $b$, $u$ \\
\hline
UV & {\it Swift}/UVOT & $uvw1$, $uvm2$, $uvw2$  \\
\hline
X-ray & {\it Swift}/XRT & 0.3--10 keV \\
& {\it Swift}/BAT & 15--50 keV \\
\hline
Gamma-ray & AGILE GRID & 100 MeV -- 30 GeV \\
\hline \hline
\end{tabular}
\caption{Observatories contributing to the presented data set of PKS 1510$-$089 at different frequencies.}
\end{center}
\end{table}

\section{AGILE observations: data analysis and results}
 
The AGILE scientific instrument (Tavani et al.~2009) is very compact and combines four detectors that provide broad band coverage from hard X-rays to
$\gamma$-rays: a Silicon Tracker (ST) optimized for $\gamma$-ray imaging in
the 30 MeV -- 30 GeV energy range (Prest et al.~2003; Barbiellini et
al.~2001); a co-aligned coded-mask X-ray imager sensitive in the 18--60 keV
energy range (SuperAGILE; Feroci et al.~2007); a non-imaging CsI(Tl)
Mini-Calorimiter sensitive in the 0.3--100 MeV energy range (MCAL; Labanti et
al.~2009); and a segmented anticoincidence system (ACS; Perotti et
al.~2006). The combination of ST, MCAL and ACS forms the GRID that assures the
$\gamma$-ray detection. 

The AGILE satellite observed PKS 1510$-$089 between 1 March 2009 00:01 UT and
31 March 2009 11:41 UT (JD 2454891.5-2454921.0), for a total of 325 hours of
effective exposure time, during which the source moved from $\sim 25^\circ$ to
$\sim 55^\circ$ off the AGILE pointing direction.

AGILE-GRID data were analyzed, starting from the Level--1 data, using the AGILE Standard Analysis Pipeline and the AGILE Scientific Analysis
Package. Counts, exposure, and Galactic background $\gamma$-ray maps, the latter based on the diffuse emission model developed for AGILE (Giuliani
et al.~2004), were generated with a bin size of $0.25^{\circ} \times 0.25^{\circ}$ for photons with energies E $>$ 100 MeV. We selected only the events
flagged as confirmed $\gamma$-ray events, and not collected during the South Atlantic Anomaly or whose reconstructed directions form angles with
the satellite-Earth vector smaller than 80$^{\circ}$, in order to reduce the $\gamma$-ray Earth albedo contamination. 

\begin{figure*}[!hhht]
\centering
\includegraphics[width=14cm]{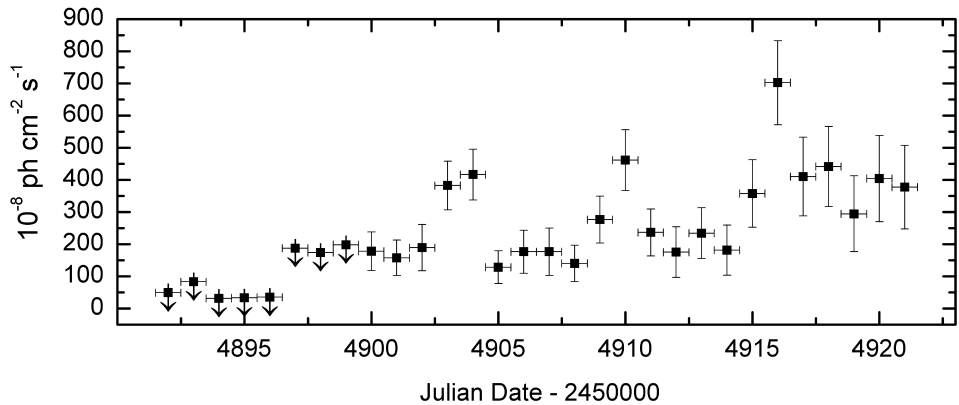}
\caption{AGILE $\gamma$-ray light curve of PKS 1510--089 between 1 and 30
  March 2009 (JD 2454891.5--2454921.5) for E $>$ 100 MeV at 1-day
  resolution. The downward arrows represent 2-$\sigma$ upper limits.} 
\label{AGILE2009}
\end{figure*}

PKS 1510$-$089 was detected over the period 1--30 March 2009 at a significance
  level of 19.9-$\sigma$, with an average $\gamma$-ray flux of $F_{\rm
  E\,>\,100\, \rm MeV}$ = (162 $\pm$ 12) $\times$ 10$^{-8}$ photons cm$^{-2}$
  s$^{-1}$. Figure~\ref{AGILE2009} shows the $\gamma$-ray light curve of March
  2009 with 1-day resolution for photons with energy higher than 100 MeV. The
  average $\gamma$-ray flux as well as the daily values were derived by the
  maximum likelihood analysis, according to the procedure described in Mattox
  et al.~(1993): first, the entire period was analyzed to determine the
  diffuse emission parameters, then the source flux density was estimated
  independently for each of the 1-day periods with the diffuse parameters
  fixed at the value obtained in the first step. At the beginning of the
  observation period the source was not so active in $\gamma$-rays as
  later. AGILE-GRID does not detect PKS 1510$-$089 at a significance level
  higher than 3-$\sigma$ between 1 and 8 March 2009 and only upper limits at
  95$\%$ confidence level are obtained\footnote{When the significance level of
  the detection obtained by the maximum likelihood analysis is $<$ 3-$\sigma$
  an upper limit at 95$\%$ confidence level is calculated (see Mattox et
  al.~1996).}. Instead, in the period 9--30 March 2009 (JD
  2454899.5--2454921.5) different flaring episodes were detected and
  considering only this time interval the significance of detection slightly
  increases to 21.5-$\sigma$, with an average flux of (311 $\pm$ 21) $\times$
  10$^{-8}$ photons cm$^{-2}$ s$^{-1}$. 

In the period 9--30 March 2009 the AGILE 95$\%$ maximum likelihood contour
level barycenter of the source is l = 351.278$^{\circ}$ b = 40.058$^{\circ}$,
with a distance between the $\gamma$-ray position and the radio position (l =
351.289$^{\circ}$, b = 40.139$^{\circ}$) of 0.08$^\circ$ and an overall AGILE
error circle of radius r = 0.15$^{\circ}$.

The $\gamma$-ray light curve shows a complex structure with outbursts having a
duration of 4--5 days and approximately symmetrical profile that could be an
indication that the relevant timescale is the light crossing time of the
emitting region. The peak level of activity with daily integration was $F_{\rm
  E\,>\,100\, \rm MeV}$ = (702 $\pm$ 131) $\times$ 10$^{-8}$ photons cm$^{-2}$
s$^{-1}$ on 25 March 2009 (JD 2454916.0)\footnote{Integrating between 0:00 UT
  and 23:59 UT for each day.}. This is the highest $\gamma$-ray flux observed
from this source on daily timescale and one of the highest fluxes detected
from a blazar. We note that the increasing of the error on the flux estimation
throughout the whole observation period is related to the fact that with the
increase of the off-axis angle between the center of the field of view of the
GRID and the position of the source, the possible background contamination and
the GRID calibration uncertainties become slightly higher. Despite the
different energy range and T$_{start}$ of the daily light curves, the AGILE results are in agreement with the {\it Fermi}-LAT results presented in Abdo et al.~(2010a).

Unfortunately, the source was located substantially off-axis in the SuperAGILE
field of view during the whole observation period, thus preventing us to
extract a constraining upper limit in the 18--60 keV energy band with SuperAGILE.

The $\gamma$-ray spectrum during the first period of high activity detected by
AGILE (9--16 March 2009; JD 2454899.5--2454907.5) can be fitted with a power
law of photon index $\Gamma$ = 1.95 $\pm$ 0.15. The photon index is calculated
with the weighted least squares method by considering four energy bins:
100--200 MeV, 200--400 MeV, 400 MeV--1 GeV, and 1--3 GeV. A similar value of
the photon index ($\Gamma$ = 1.96 $\pm$ 0.24) is estimated also in the period
of the highest activity (23--27 March 2009; JD
2454913.5--2454918.5). Considering that during this second period the source
was far off-axis with respect to the center of the GRID field of view, the
energy bin 1--3 GeV is not considered in order to reduce the possible
background contamination for high off-axis angles at these energies. Comparing
with the average photon index measured by EGRET ($\Gamma$ = 2.47 $\pm$ 0.21,
Hartman et al.~1999) and that measured by {\it Fermi}-LAT over the first
eleven months of observations ($\Gamma$ = 2.41 $\pm$ 0.01; Abdo et al.~2010b),
the values observed by AGILE confirm a hardening of the $\gamma$-ray spectrum
during the very high activity detected in March 2009. A similar hard
$\gamma$-ray spectrum of PKS 1510$-$089 was observed by AGILE during the flaring episode of March 2008 (D'Ammando et al.~2009a). 
  
The correlation between the flux level and the spectral slope in the
$\gamma$-ray energy band was extensively studied by means of the analysis of
all the EGRET data, but a firm result for the blazars is not found
(Nandikotkur et al.~2007). A `harder-when-brighter' behaviour in $\gamma$ rays
of PKS 1510$-$089 was detected by {\it Fermi}-LAT over a long timescale
(August 2008--June 2009) considering only flux levels of F (E $>$ 200 MeV)
$\gtrsim$ 2.4 $\times$ 10$^{-7}$ photons cm$^{-2}$ s$^{-1}$ (Abdo et
al. 2010a). A similar trend was observed in the EGRET {\it era} only for 3C
279 (Hartman et al.~2001) and marginally for PKS 0528$+$134 (Mukherjee et
al.~1996), but this could be due to the fact that these were the only objects
for which a long-term monitoring in $\gamma$-rays was performed by EGRET. At
present, with AGILE and {\it Fermi} we are able to follow a large number of
blazars on very long timescales and investigate this behaviour on a larger
sample of objects. In this context, considering the long period of high
activity recently observed in $\gamma$-rays, PKS 1510$-$089 is one of the best test cases, together with 3C 454.3 (see e.g. Vercellone et al.~2010).

\section{{\it Swift} observations: data analysis and results}

The {\it Swift} satellite (Gehrels et al.~2004) performed 15 ToOs on PKS
1510$-$089 between 11 and 30 March 2009, triggered by the high $\gamma$-ray
activity of the source. The observations were performed with all the three
onboard instruments: the X-ray Telescope (XRT; Burrows et al.~2005, 0.2--10.0
keV), the UltraViolet Optical Telescope (UVOT; Roming et al.~2005, 170-600
nm), and the Burst Alert Telescope (BAT; Barthelmy et al.~2005, 15--150 keV).

\subsection{{\it Swift}/BAT data}

As part of its normal operations, BAT collects data over a wide area of the sky in its survey mode. The survey data in the 15--50 keV band is 
used to produce sky images in which hard X-ray sources can be detected using
the standard {\it Swift} analysis software, following the procedures described
in Krimm et al.~(2008, and the reference therein; see also\footnote{http://swift.gsfc.nasa.gov/docs/swift/results/transients}.). 

During March 2009, we detected two short flaring episodes from PKS 1510$-$089. The first covered approximately 2 days beginning on 8 March 2009
(JD 2454899.0), with an average count rate of (0.006 $\pm$ 0.002) counts
s$^{-1}$ cm$^{-2}$ (15--50 keV), which corresponds to 28 mCrab and peaking on
9 March 2009 at 40 mCrab (Krimm et al.~2009). A second weaker episode occurred
on 29 March 2009 (JD 2454920.0), where the average count rate was (0.003 $\pm$
0.001) counts s$^{-1}$ cm$^{-2}$, corresponding to 15 mCrab. 

As a comparison we report that during the period before the March 2008
$\gamma$-ray flare detected by AGILE (D'Ammando et al.~2009a), {\it Swift}/BAT
recorded flaring episodes from PKS 1510$-$089 on 16 February 2008 (JD
2454513.0), 21 February 2008 (JD 2454518.0) and 24 February 2008 (JD
2454521.0), with count rates of approximately (0.005 $\pm$ 0.002) counts s$^{-1}$ cm$^{-2}$ on each of those observations.

\subsection{{\it Swift}/XRT data}

The XRT data were processed with standard procedures ({\tt xrtpipeline}
v0.12.4), filtering, and screening criteria by using the {\tt Heasoft} package
(v.6.8). The source count rate was low during the whole campaign (count rate
$<0.5$ counts s$^{-1}$), so we only considered photon counting (PC) data and
further selected XRT event grades 0--12. Pile-up correction was not
required. Source events were extracted from a circular region with a radius
between 15 and 20 pixels (1 pixel $\sim 2.36\arcsec$), depending on the source
count rate, while background events were extracted from a circular region with
radius 40 pixels away from background sources. Ancillary response files were
generated with {\tt xrtmkarf}, and account for different extraction regions,
vignetting and PSF corrections. We used the spectral redistribution matrices
v011 in the Calibration Database maintained by HEASARC. Considering the very
low number of photons collected on 26 and 27 March we summed the data of these
two consecutive days in order to have enough statistics to obtain a good spectral fit.

All spectra were rebinned with a minimum of 20 counts per energy bin to allow
$\chi^2$ fitting within {\sc XSPEC} (v12.5.1; Arnaud 1996). We fit the
individual spectra with a simple absorbed power law, with a neutral hydrogen
column fixed to its Galactic value ($6.89\times10^{20}$ cm$^{-2}$; Kalberla et al.~2005).  
The fit results are reported in Table~\ref{XRT_March2009}. The comparison between our results and those presented by Abdo et al.~(2010a) reveals some
  differences, which we ascribe to their using a lower number of counts per bin than generally adopted to allow $\chi^{2}$ fitting ($\geq$ 20
  counts/bin). Indeed, these differences increase when the number of counts decreases.

\begin{table*}[th!]
\caption{Log and fitting results of {\it Swift}/XRT observations of PKS 1510$-$089 during March 2009. Power law model with $N_{\rm
H}$ fixed to Galactic absorption (6.89 $\times$ 10$^{20}$ cm$^{-3}$) is used. $^{a}$ Unabsorbed flux.}
\centering
\begin{tabular}{cccccc}
\hline
\hline
\noalign{\smallskip}
\multicolumn{1}{c}{Start Time} &
\multicolumn{1}{c}{Stop Time} &
\multicolumn{1}{c}{Exp. Time} &
\multicolumn{1}{c}{Photon Index} &
\multicolumn{1}{c}{Flux 0.3--10.0 keV$^{a}$} &
\multicolumn{1}{c}{$ \chi^{2}_{\rm red}$ (d.o.f.)} \\
\multicolumn{1}{c}{(yyyy-mm-dd hh:mm:ss)} &
\multicolumn{1}{c}{(yyyy-mm-dd hh:mm:ss)} &
\multicolumn{1}{c}{ (sec) } &
\multicolumn{1}{c}{$\Gamma$}&
\multicolumn{1}{c}{($\times$ 10$^{-12}$ erg cm$^{-2}$ s$^{-1}$}) &
\multicolumn{1}{c}{} \\
\hline
\noalign{\smallskip}
2009-03-11 15:07:49 & 2009-03-11 23:25:56 & 4891 & $1.54 \pm 0.10$ & $9.28 \pm 0.77$ & 0.919 (35) \\
2009-03-12 15:13:44 & 2009-03-12 20:29:50 & 4842 & $1.45 \pm 0.09$ & $10.86 \pm 0.84$ & 0.867 (38)\\
2009-03-17 04:29:17 & 2009-03-17 11:07:56 & 4869 & $1.51 \pm 0.11$ & $8.67 \pm 0.77$ & 1.041 (31) \\
2009-03-18 00:02:10 & 2009-03-18 08:03:56 & 4777 & $1.37 \pm 0.10$ & $9.37 \pm 0.88$ & 1.009 (27) \\
2009-03-19 19:14:17 & 2009-03-19 22:39:56 & 2501 & $1.61 \pm 0.14$ & $8.82 \pm 0.99$ & 1.074 (15) \\
2009-03-20 22:34:29 & 2009-03-21 02:00:58 & 2010 & $1.28 \pm 0.18$ & $10.03 \pm 1.33$ & 0.884 (11) \\
2009-03-22 01:30:19 & 2009-03-22 06:38:57 & 2242 & $1.43 \pm 0.18$ & $9.51 \pm 1.38$ & 0.843 (12) \\
2009-03-22 03:05:15 & 2009-03-22 03:48:27 & 2580 & $1.50 \pm 0.17$ & $8.59 \pm 1.14$ & 0.913 (13) \\
2009-03-23 11:25:24 & 2009-03-23 16:32:57 & 2640 & $1.64 \pm 0.15$ & $7.77 \pm 0.91$ & 0.705 (15) \\
2009-03-24 03:27:22 & 2009-03-24 22:57:57 & 1972 & $1.60 \pm 0.21$ & $7.97 \pm 1.26$ & 0.768 (10) \\
2009-03-25 13:23:14 & 2009-03-25 22:58:57 & 2447 & $1.53 \pm 0.14$ & $9.50 \pm 1.13$ & 1.102 (15) \\
2009-03-26 16:27:40 & 2009-03-27 08:43:56 & 5003 & $1.50 \pm 0.13$ & $8.39 \pm 0.89$ & 0.871 (22) \\
2009-03-28 05:18:45 & 2009-03-28 08:48:58 & 2657 & $1.31 \pm 0.14$ & $10.16 \pm 1.30$ & 0.997 (17) \\
2009-03-30 10:18:41 & 2009-03-30 12:12:57 & 2544 & $1.40 \pm 0.14$ & $9.93 \pm 1.25$ & 1.230 (16) \\

\noalign{\smallskip}
\hline
\hline
\noalign{\smallskip}
\end{tabular}
\\
\label{XRT_March2009}
\end{table*}

\begin{figure}[!t]
\epsfig{file=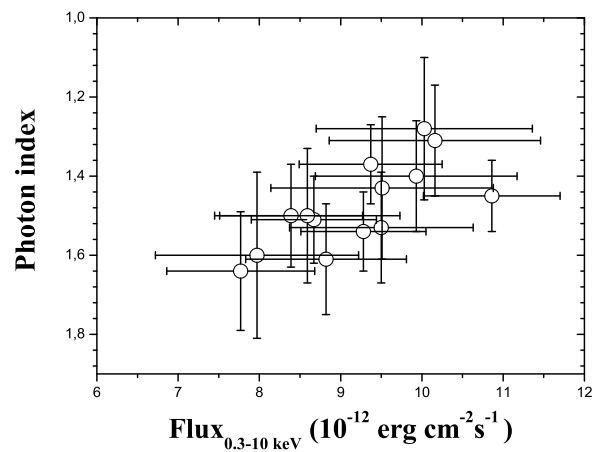, width=90mm}
\caption{{\it Swift}/XRT photon index as a function of the 0.3--10 keV flux.} 
\label{XRT:hwb}
\end{figure}

During the 15 ToOs performed in March 2009, {\it Swift}/XRT observed the
source with a 0.3--10.0 keV flux in the range (0.8 -- 1.1) $\times$ 10$^{-11}$
erg cm$^{-2}$ s$^{-1}$, a somewhat fainter state with respect to the high flux
level observed in August 2006 (Flux$_{\rm 0.5-10\,keV}$ = (1.1 -- 1.8)
$\times$ 10$^{-11}$ erg cm$^{-2}$ s$^{-1}$). Instead, a hint of spectral trend
seems to be present in the X-ray data. Figure~\ref{XRT:hwb} shows the XRT
photon indexes as a function of the X-ray fluxes in the 0.3--10 keV: a possible hardening of the spectrum with the increase of the flux is observed, in
spite of the large errors. A similar harder-when-brighter behaviour in X-rays
was observed for PKS 1510$-$089 in August 2006 (Kataoka et al.~2008){\bf ,}
March 2008 (D'Ammando et al.~2009a), and over the period January--June 2009 (Abdo et al.~2010a). 

The significant spectral evolution observed by XRT, with the photon index
changing from 1.3 to 1.6, is not usual for FSRQs, for which only little X-ray
variability is detected both on short and long timescales. Moreover, the
average X-ray photon index observed over March 2009 for this object
($<\Gamma>$ = 1.48 $\pm$ 0.03) is lower than that of the radio-loud quasars ($<\Gamma>$ = 1.66 $\pm$ 0.07; Lawson et al.~1992, Cappi et al.~1997) and is
more similar to that of the high-redshift (z $>$ 2) quasars (e.g. Page et
al.~2005). Since usually in FSRQs the X-ray energy range samples the
low-energy tail of the external Compton (EC) component, as already discussed
in Kataoka et al.~(2008) and Abdo et al.~(2010a), such hard spectral indexes
in X-rays should imply a very flat energy distribution, challenging the
standard shock models of particle acceleration. In a standard shock model
these hard X-ray spectra produce relativistic electrons with an energy
spectrum harder than the canonical power law distribution N($\gamma$)
$\propto$ $\gamma$$^{-2}$, or alternatively, as discussed in Sikora et
al.~(2002), another mechanism (e.g.~instabilities driven by shock-reflected
ions, Hoshino et al.~2002; or magnetic reconnection, Romanova $\&$ Lovelace
1992) should energize the low energies electrons that typically produce the X-ray emission in the EC model. 

\subsection{{\it Swift}/UVOT data}

\begin{figure*}[!hhht]
\sidecaption
\includegraphics[width=11.0cm]{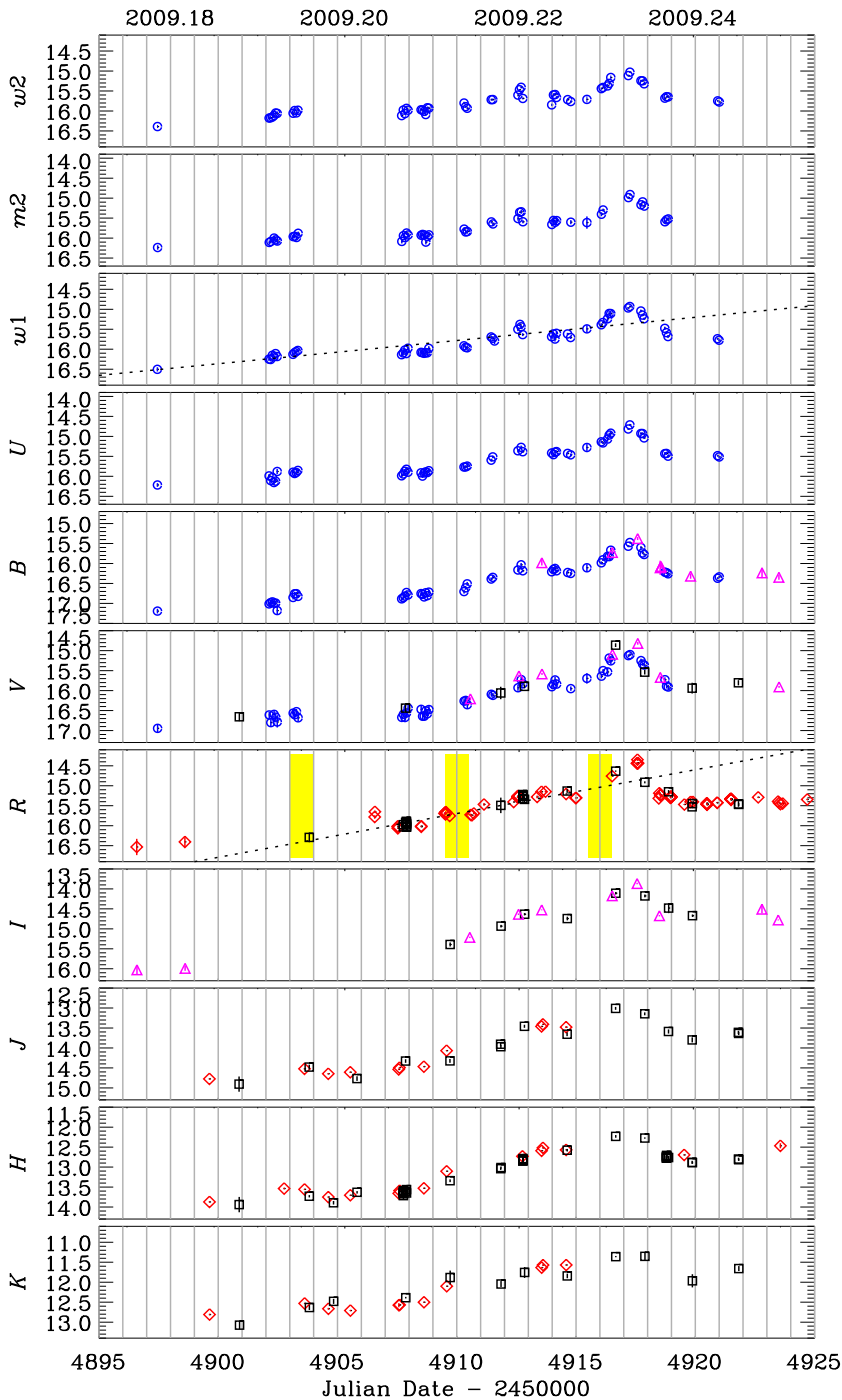}
\caption{Light curves collected in near-IR, optical and UV bands, between 5
  March and 2 April 2009 (JD 2454895.5--2454924.5). Blue circles represent the
  UVOT data in $v$, $b$, $u$, $uvw1$, $uvm2$, $uvw2$ filters. Red diamonds
  represent GASP data in $R$, $J$, $H$, and $K$ bands. Magenta triangles
  represent WEBT data in $I$, $V$, and $B$ \,bands. REM data in $V$, $R$, $I$,
  $J$, $H$, and $K$ bands are represented with black squares. Yellow regions
  in $R$-band light curve indicate the peaks of $\gamma$-ray activity observed by AGILE.} 
\label{WEBT2009}
\end{figure*}

UVOT observed PKS 1510$-$089 in all its optical ($v$, $b$, and $u$) and UV
($uvw1$, $uvm2$, and $uvw2$) photometric bands. Data were reduced with the
{\tt uvotmaghist} task of the HEASOFT package. Source counts were extracted
from a circular region of 5 arcsec radius, centred on the source, while the
background was estimated from a surrounding annulus with 8 and 18 arcsec
radii. In Fig.~\ref{WEBT2009} UVOT magnitudes are displayed with blue circles.

\section{GASP observations in radio, near-IR and optical bands: data analysis and results}

The GLAST-AGILE Support Program (Villata et al.~2008, 2009a) is a project born
from the Whole Earth Blazar Telescope\footnote{\tt
  http://www.oato.inaf.it/blazars/webt.} in 2007. It is aimed at providing long-term monitoring in the optical ($R$-band), near-IR, and radio bands of 28
$\gamma$-ray-loud blazars during the lifetime of the two $\gamma$-ray satellites.

The $R$-band GASP observations of PKS 1510$-$089 in the period considered in
this paper were performed by the following observatories: Abastumani, Calar
Alto\footnote{Calar Alto data were acquired as part of the MAPCAT project
  http://www.iaa.es/~iagudo/research/MAPCAT.}, Castelgrande, L'Ampolla, La
Silla (MPG/ESO), Lulin (SLT), Roque de los Muchachos (KVA and Liverpool), San
Pedro Martir, St. Petersburg, and Valle d'Aosta. These data were calibrated
with respect to stars 2--6 by Raiteri et al.~(1998). The GASP observation in
$R$-band showed that after a low intensity period in February 2009, with the
source observed constantly at $\sim$ 16.5 mag, the optical activity of the
source greatly increased in March 2009 (Villata et al.~2009b) with an intense
flare on 27 March (JD $\sim$ 2454917.6), after a brightening by almost 1 mag
in 3 days, reaching $R$ = 14.35 $\pm$ 0.03 mag (Larionov et al.~2009; see
Fig.~3 and Fig.~4). Near-IR data in $J, H$, and $K$ \,bands are from Campo
Imperatore and Roque de los Muchachos (Liverpool), whereas WEBT data in $B$,
$V$, and $I$ \,bands were taken at Castelgrande, La Silla and
St.~Petersburg. Data collected by GASP and WEBT observatories during March
2009 are reported in Fig.~\ref{WEBT2009} with red diamonds and magenta triangles, respectively. 

\begin{figure}[!hhh]
\centering
\includegraphics[width=9.0cm]{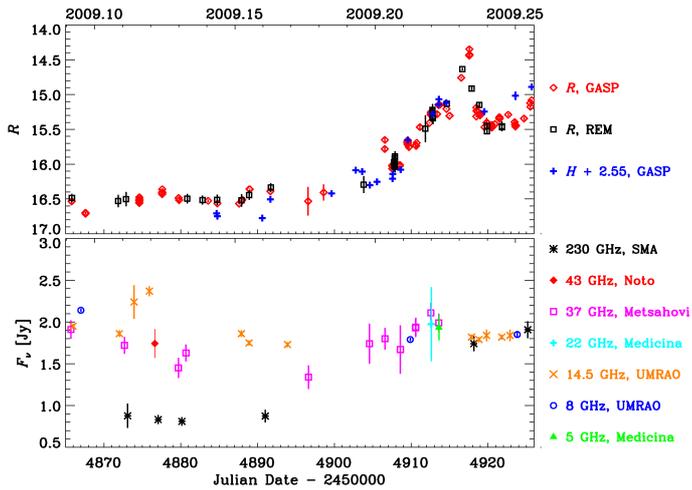}
\caption{$R$-band light curve of PKS 1510$-$089 obtained by GASP and REM
  during February--March 2009, together with the $H$ band data by GASP (top panel), compared with the radio flux densities at different frequencies (bottom panel).} 
\label{radio}
\end{figure}

Radio fluxes were measured at: Submillimeter Array\footnote{These data were obtained as part of the normal monitoring program initiated by the SMA (see
  Gurwell et al.~2007).} (SMA, 230 GHz), Medicina (22 and 5 GHz), Mets\"ahovi
  (37 GHz), Noto (43 GHz), and UMRAO (14.5 and 8.0 GHz). The radio data are
  shown in Fig.~\ref{radio}, together with the optical ($R$-band) and near-IR ($H$-band) light curve of PKS 1510$-$089 in February--March 2009. 

We  point out that in those cases where different datasets were present in the
  same band, we performed a careful data analysis to determine possible
  offsets and corrected for them to get homogeneous light curves, as usually
  done for GASP--WEBT data (see e.g.~Villata et al.~2002, Raiteri et
  al.~2005).

\section{REM observations: data analysis and results}

During March 2009 photometric near-IR and optical observations were carried
out with REM (Zerbi et al.~2004), a robotic telescope located at the ESO Cerro
La Silla observatory (Chile). The REM telescope has a Ritchey-Chretien
configuration with a 60 cm f/2.2 primary and an overall f/8 focal ratio in a
fast moving alt-azimuth mount providing two stable Nasmyth focal stations. At
one of the two foci, the telescope simultaneously feeds, by means of a
dichroic, two cameras: REMIR for the near-IR (Conconi et al.~2004) and ROSS
(Tosti et al.~2004) for the optical. Both cameras have a field of view of 10
$\times$ 10 arcmin and imaging capabilities with the usual near-IR (z, $J$,
$H$, and $K$) and Johnson-Cousins $VRI$ filters. The REM software system
(Covino et al.~2004) is able to manage complex observational strategies in a
fully autonomous way. All raw near-IR/optical frames obtained with the REM
telescope were then corrected for dark, bias, and flat field following
standard recipes. Instrumental magnitudes were obtained via aperture
photometry and absolute calibration has been performed by means of secondary
standard stars in the field (see Raiteri et al.~1998; Smith and Balonek
1998). The data presented here were obtained by GO program for AOT19 (PI:
F.~D'Ammando). REM data in $V$, $R$, $I$, $J$, $H$, and $K$ \,bands are
reported in Fig.~\ref{WEBT2009} with black squares. 

\section{VLBA observations: data analysis and results}

\begin{figure}[!hhht]
\center
\includegraphics[width=9cm]{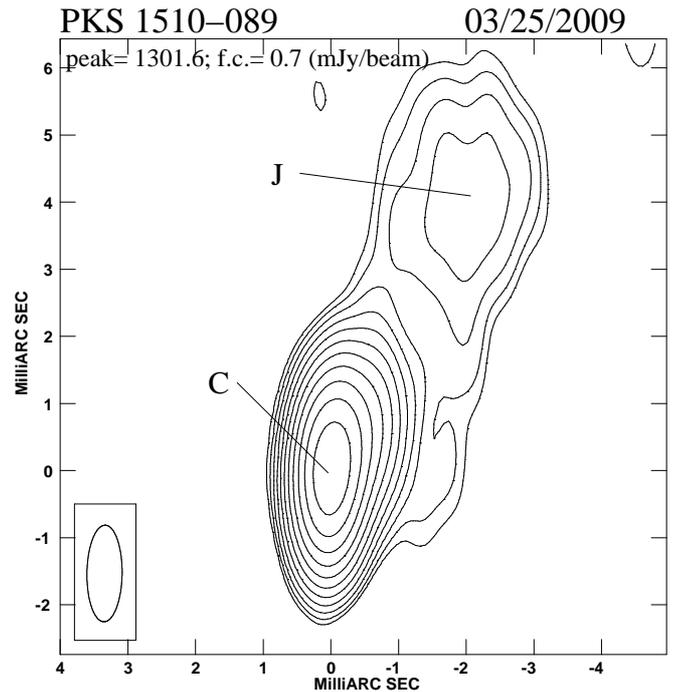}
\caption{VLBA image at 15 GHz of PKS 1510$-$089 on 25 March 2009. On the image
  we provide the restoring  beam, plotted in the bottom left corner, the peak flux density in mJy beam$^{-1}$, and the first contour (f.c.) intensity in mJy beam$^{-1}$, which is 3 times the off-source noise level. Contour levels increase by a factor of 2.}
\label{VLBA}
\end{figure}

When observed with the high spatial resolution provided by the Very Long
Baseline Interferometer (VLBI) technique, PKS 1510$-$089 shows a radio
structure dominated by the core region from which the pc-scale jet emerges
forming an angle of $-$28$^{\circ}$, i.e. in the north-west
direction. Monitoring campaigns pointed out the ejection of new superluminal components, and followed their evolution throughout several observing epochs
(see Marscher et al.~2010). \\
As part of the MOJAVE program (Lister et al.~2009), PKS 1510$-$089 is
frequently observed with the VLBA at 15 GHz (see Fig.~\ref{VLBA}). To study
variations in the source structure, we retrived the {\it uv}-datasets and we
imported them into NRAO AIPS package. Final images both in total intensity and
polarization have been produced after a few phase-only self-calibration
iterations (for more details on calibration and image analysis see Orienti et
al.~2010). From the analysis of the multi-epoch observations obtained between
2007 and 2010 it was possible to follow the separation between the core
region, considered stationary, and three emitted knots. From linear regression
fits we obtain highly superluminal apparent separation velocities between
16$c$ and 20$c$, consistent with the separation speed derived for other knots
(e.g.~Homan et al.~2002), and at different frequency (Marscher et al.~2010),
indicating an intrinsic separation velocity $\beta_{\rm intr} >0.9989$ and an orientation with the line of sight $\theta < 5^{\circ}$.\\
From the pc-scale resolution images we could separate the core flux density
from the jet emission, in order to study the lightcurve and polarization
properties of each of them. Between 2007 and 2010, three main episodes of
enhanced radio luminosity have been detected, successively followed by marked
flux density decreases. By contrast, no obvious trend of the polarization
percentage has been found.\\
After March 2009, the flux density of the core region, that was in a minimum,
started to increase, while the flux density of the jet component did not show
any significant luminosity changes. A remarkable aspect observed just after
the March 2009 flare is the abrupt change of about 75$^{\circ}$ in the
polarization angle of the core, whereas the fractional polarization remains
constant. More details on the long-term radio monitoring of this source are presented in Orienti et al.~(2010).\\ 

\section{Discussion}

\subsection{Thermal signatures in the radio-to-UV spectrum}

Evidence of thermal emission in blazars has been detected in a limited number
of objects until now, and usually observed during low activity states (e.g.~3C
279, Pian et al.~1999; 3C 273, Grandi $\&$ Palumbo 2004; 3C 454.3, Raiteri et
al.~2007; AO 0235$+$164, Raiteri et al.~2008; NRAO 150, Acosta-Pulido et
al.~2010). This is because, contrary to the other AGNs, in blazars the jet
emission at small angles with respect to the line of sight of the observer is strongly amplified due to the relativistic beaming effect, overwhelming all
the thermal contributions. However, these thermal features are less prominent
in blazars but not absent, at least for FSRQs, which present a radiatively
efficient accretion disk and broad emission lines. 

The knowledge of these thermal features could have important consequences not
only for the low-energy part of the spectrum, in which the thermal emission
could be directly observed, but also for the high-energy part of the spectrum,
because the photons produced by the accretion disk, directly or through the
reprocessing by the broad line region (BLR) or the dusty torus, can become the
main source of seed photons for the EC mechanism that usually dominates the $\gamma$-ray emission of FSRQs (Ghisellini et al.~1998). 

In this context PKS 1510$-$089 seems to be peculiar. Since the synchrotron
peak of this source is usually observed in mid-infrared band (see Bach et
al.~2007; Nieppola et al.~2008), the optical/UV part of the spectrum is not
dominated by the non-thermal jet emission, allowing us to observe directly the thermal manifestation of both the accretion disk (the so-called ``big blue
bump'', e.g.~Laor 1990) and of the BLR (the so-called ``little blue bump'',
e.g.~Wills et al.~1985) also during high activity states, like in mid-March
2008 (see D'Ammando et al.~2009a). Moreover, an excess at far-IR wavelengths
was observed in PKS 1510$-$089 by IRAS (Tanner et al.~1996), likely due to
dust radiation from the torus. Recently, {\it Spitzer} observations of PKS
1510$-$089 produced only an upper limit of 2 $\times$ 10$^{45}$ erg s$^{-1}$
to the thermal emission from dust torus (Malmrose et al.~2011).

In order to distinguish between non-thermal and thermal emission
contributions, we built spectral energy distributions (SEDs) from radio to UV
with GASP-WEBT, REM and {\it Swift}/UVOT data. Conversion of magnitudes into
de-reddened flux densities was obtained by adopting the Galactic absorption
value $A_{B}$ = 0.416 from Schlegel et al.~(1998), the extinction laws by
Cardelli et al.~(1989) and the magnitude-flux calibrations by Bessell et al.~(1998). 

As for the {\it Swift}/UVOT data, we noticed that PKS 1510$-$089 has a {\it
  b-v} $\sim$ 0.3 that is out of the validity range indicated by Poole et
  al.~(2008) for their flux calibrations in the UV bands. Hence, following
  Raiteri et al.~(2010, 2011) we calculated effective wavelengths
  $\lambda_{\rm eff}$, count rate to flux density conversion factors $\rm CF$,
  and amount of Galactic extinction $A_\Lambda$ for each UVOT band, by folding
  the quantities of interest with the source spectrum and effective areas of UVOT filters. The results are shown in Table 3.

 \begin{table}
\caption{Results of the UVOT calibration procedure for PKS 1510-089.}
\label{caluvot}     \centering                         \begin{tabular}{l c c c}     
\hline\hline                Filter & $\lambda_{\rm eff}$ & $\rm CF$ & $A_\Lambda$ \\     
   & \AA\ & $10^{-16} \rm \, erg \, cm^{-2} \, s^{-1} \, \AA^{-1}$ & mag \\
\hline                         $v$    & 5422 & 2.61 & 0.33\\       $b$    & 4346 & 1.47 
& 0.43\\
  $u$    & 3466 & 1.65  & 0.52\\
  $uvw1$ & 2633 & 4.21  & 0.73\\
  $uvm2$ & 2251 & 8.45  & 0.90\\
  $uvw2$ & 2059 & 6.31  & 0.86\\
\hline\hline                                 \end{tabular}
\end{table}

\begin{figure}[!hhhb]
\centering
\includegraphics[width=9.1cm]{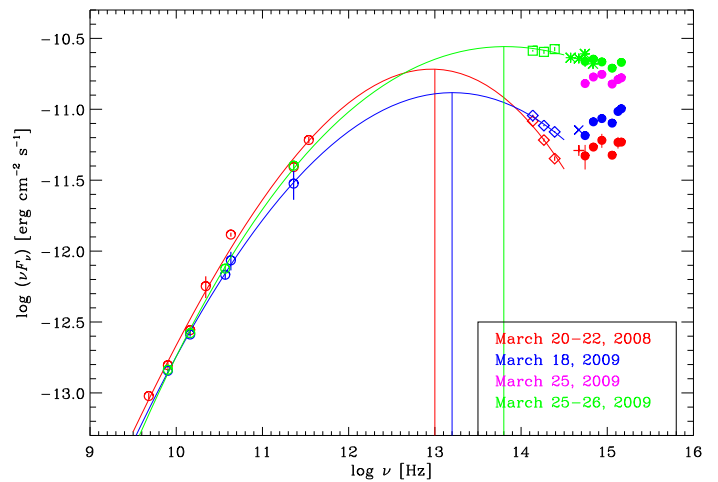}
\caption{SEDs of the low-energy part of the spectrum constructed with data
  collected by GASP-WEBT (empty circle), KVA (plus sign), Abastumani (cross),
  St.~Petersburg (asteriks), Campo Imperatore (diamonds), REM (squares), and
  {\it Swift}/UVOT (filled circles) during March 2008 and March 2009. A cubic
  polynomial fit was applied to the SEDs for locating the position of the synchrotron peak.} 
\label{SEDoptical}
\end{figure}

UVOT count rates were thus converted into flux densities and then corrected
for Galactic extinction adopting the $\rm CF$ and $A_\Lambda$ reported in
Table 3. In Figure 6 we display three SEDs of March 2009 as well as the March
2008 SED already published by D'Ammando et al.~(2009a). The comparison between
the March 2008 SED obtained with standard calibrations (Figure 3 of D'Ammando
et al.~2009a) and the same SED obtained by applying the procedure described above reveals that the dip in the $uvw1$ band remains, though it is now less
pronounced. This means that it is not due to calibration problems, but likely
reflects a true intrinsic feature. Indeed, we expect the little blue bump due
to BLR emission (MgII, FeII, and Balmer lines) to peak in the $u$--$b$
frequency range. Near-IR to UV spectra of PKS 1510$-$089 in March 2009 have
also been presented by Abdo et al.~(2010a), but for a different choice of
epochs, which prevents a detailed comparison with our results. However, as
expected, we notice a deeper $w1$ dip than that we obtained with our more accurate calibrations. 

In other FSRQs, such as 3C 454.3, the presence of the little and big blue
bumps was detected only during low activity state of the source (see Raiteri
et al.~2008, Vercellone et al.~2010). Instead, the SED of PKS 1510$-$089
collected on 18 March 2009 confirmed the evidence of thermal signatures in the
optical/UV spectrum of the source also during high $\gamma$-ray states, with a
contribution in the optical part likely due to the little blue bump and a
significant rise of the spectrum at UV due to the accretion disk emission, as
already observed on 20--22 March 2008. On the other hand, the broad band
spectrum of PKS 1510$-$089 from radio-to-UV during 25--26 March 2009 shows a
flat spectrum in the optical/UV energy band, suggesting an important
contribution of the synchrotron emission in this part of the spectrum during
the brightest $\gamma$-ray flaring episode and therefore a significant shift
of the synchrotron peak, usually observed in this source in the infrared
band. We derived an estimate of the frequency of the synchrotron peak in the
three SEDs applying a cubic polynomial fit (see e.g.~Kubo et al.~1998) to the
radio and IR data, which are likely due to pure synchrotron, as shown in
Fig.~\ref{SEDoptical}. The synchrotron peak shifted from $\nu$ = 1.5 $\times$
10$^{13}$ Hz to $\nu$ = 6.5 $\times$ 10$^{13}$ Hz between 18 and 26 March
2009. Similarly to the harder-when-brighter behaviour observed in X-rays, this
is a typical behaviour of high-frequency-peaked BL Lacs (HBLs) and not so
commonly observed in FSRQs such as PKS 1510$-$089, even if this could be
partially due to the fact that it is more difficult to obtain a long term
monitoring of the synchrotron peak of FSRQs in the IR band with respect to the optical/UV and X-ray band. 

Abdo et al.~(2010a) analyzing the UVOT data of PKS 1510$-$089 with standard
  calibrations (and thus with possible biases in the results, see above) found
  an anticorrelation between UV flux and UV hardness ratio in the period
  January-June 2009, suggesting a different level of contamination in UV by
  the high-energy branch of synchrotron emission depending on the activity
  level. This is in agreement with our results, even if the importance of the
  increase of the synchrotron emission observed at the end of March 2009 seems
  to be much higher with respect to the general trend.    

The UVOT data reported in the SED of 25--26 March 2009 are collected at JD =
2454916.46, instead the BVRI data are collected by St.~Petersburg at JD =
2454916.51-53, thus the separation in time between UVOT and BVRI data is about
1 hour. This rules out the possible bias related to the optical/UV variability
of the object, confirming the flat optical/UV spectrum. The NIR data for the
same SED are collected by REM at JD = 2454916.66, about 3.5 hrs after the
optical data and 4.5 hrs after the UVOT data. In this case it is not possible
to completely rule out a mismatch due to the possible rapid optical/UV
variability, but by comparison with the NIR data collected during 20-22 March
2008 and 18 March 2009, it is evident the change of the NIR spectrum, in
agreement with what we observed also in optical/UV. Moreover the comparison
between the optical/UV data collected by UVOT on 25 March 2009, during the
first UVOT exposure ($\rm JD \sim 2454916.1$, orange squares in
Fig.~\ref{SEDoptical}) and those taken about 9--10 hours after ($\rm JD \sim
2454916.5$, green triangles), reveals a noticeable spectral and flux
variation. Considering that the accretion disk is slowly variable on such
short timescales, this is another proof of the fact that the rapid increase of
the optical emission started on 25 March and peaking on 27 March is mainly due
to the synchrotron mechanism. This is also in full agreement with the
simultaneous rapid increase of the degree of optical polarization shown in
Marscher et al.~2010 (see in particular their Fig.~4). Thermal emission is
unpolarized as it reflects the random walk of atoms and ions within the
emitting region, therefore such increase of the degree of polarization is a
clear signature of a rapid and strong increase of the contribution in the optical band of a non-thermal mechanism such as the synchrotron emission. 

\subsection{Light curve behaviour and correlations}

By comparing the source behaviour observed in $\gamma$-rays by AGILE with the
ones observed from optical-to-UV by GASP-WEBT, REM and {\it Swift}/UVOT during
March 2009 we noted that while the $\gamma$-ray light curve shows evidence of
different outbursts of increasing entity, the optical and UV light curves seem
to show a gradual increase of the flux in time with a rapid flux enhancement after JD $\sim$ 2454915.0 and a single major outburst occurred between JD
2454917.25 and JD 2454917.60. A variation of 0.48 mag in 9 days (JD
2454906-2454915) was detected in the $R$-band light curve, followed by a more
rapid variation of 0.95 mag in $\sim$ 2 days. The $uvw1$ light curve shows a
similar behaviour, with a variation of 0.43 mag in 7 days and 0.78 mag in the
following 2 days. The larger increase observed in $R$-band with respect to
$uvw1$ band is in agreement with a larger contribution of the synchrotron
emission in the optical than in the UV band. A linear fit applied to the
$R$-band and $uvw1$-band light curves in Fig.~3 shows a change of slope after
JD = 2454915. The optical/UV peak seems to be delayed with respect to the
brightest $\gamma$-ray peak by 1--2 days, although we cannot exclude a very rapid optical/UV flare occurred during the gap in the optical and UV light curves before JD 2454917.25.

\begin{figure}[!hhht]
\centering
\includegraphics[width=9.0cm]{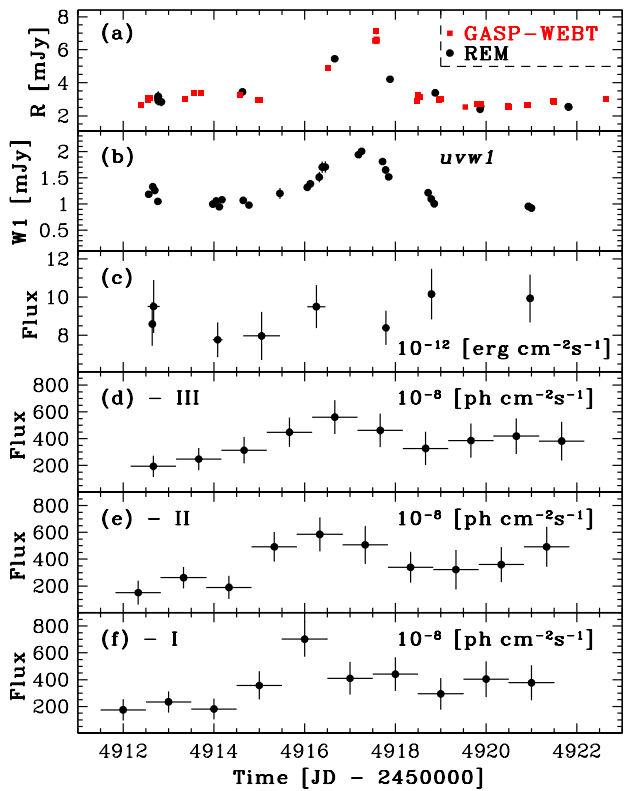}
\caption{Comparison of the $R$-band light curve of PKS 1510$-$089 (panel (a);
 GASP-WEBT: red squares, REM: black dots) with the $uvw1$ band data by {\it
 Swift}/UVOT (panel (b)), the X-ray data by {\it Swift}/XRT (panel (c)), and
 the AGILE $\gamma$-ray light curves in the three different configurations
 (panel (d), (e), (f), see Sec. 8.2 for details) during 21--31 March 2009 (JD 2454913.5--2454922.0).} 
\label{corr}
\end{figure}

Another possibility is that the delay between optical and $\gamma$-ray
emission is due to our choice of the $T_{0}$ used for building the
$\gamma$-ray daily light curve. To investigate the influence of the $T_{0}$ on
the determination of the $\gamma$-ray peak we constructed three different
light curves, considering an integration interval of 1 day and shifting the
$T_{0}$ of $\Delta$T$_1$ = 8 hours and $\Delta$T$_2$ =16 hours with respect to
our initial choice ($T_{0}$ = 0:00 UT). In Fig.~\ref{corr} the optical
($R$-band), UV ($w1$ filter) and the three $\gamma$-ray light curves built
with the different $T_{0}$ in the period 21--31 March 2009 (labelled as I, II
and III) are shown. For each $\gamma$-ray light curve we assign a weight ``1''
to the three fractional 8-hours bins that constitute the total 1-day bin with
the highest flux and a weight ``0'' to the other fractional bins. Combining
the information derived from the three light curves we can build a probability
function and estimate the fractional bin that corresponds with the highest
probability to the emission peak. In this way we estimated the $\gamma$-ray
emission peak at JD = 2454916.33 $\pm$ 1.00, leading to a reduced delay
between $\gamma$-ray and optical/UV emission, in accordance also with the
delay between the {\it Fermi}-LAT and optical light curve peaks showed in Abdo
et al.~(2010a) for the same period. Moreover, we note that a recent investigation of the ${\it Fermi}$-LAT data of PKS 1510$-$089 collected in
  the first half of 2009 by Tavecchio et al.~(2010) showed significant $\gamma$-ray variability on timescales of 6 hours and a flux peak on JD = 2454917.25, compatible with our result.     

However, a significant increase of the optical/UV emission is detected only for the brightest $\gamma$-ray flare, whereas the optical/UV activity before
JD $\sim$ 2454915 was increasing but only slowly, according to a dominant
contribution of the thermal emission, over which a subsequent rapid
variability of the synchrotron emission is superimposed (see Sec.~8.1). A progressive increase of the source activity, similar to the behaviour of the
optical light curve collected by GASP-WEBT, was observed also in the near-IR
and optical bands by the REM telescope (Fig.~\ref{WEBT2009}), suggesting that
the synchrotron is the dominant mechanism responsible for the flux enhancement
observed from near-IR to UV at the end of March 2009. In this context, the lag
of 13$\pm$1 days between the $\gamma$-ray and the optical peaks reported in
Abdo et al.~(2010a) could suggest a different origin of the dominant component
in optical band during the January and April 2009 flares with respect to the
flare occurred at the end of March 2009. This is another clue in favour of a more complex correlation between the optical and $\gamma$-ray emission in
the FSRQs, in which the optical emission could be due to the contribution of
both the synchrotron emission and the thermal accretion disk emission. Despite
the different dominant contribution in the optical part of the spectrum, a
similar photon index was observed by AGILE in the $\gamma$-ray spectrum during
the first activity period (9--16 March) and the highest activity period (23-27
March), suggesting that the same radiation mechanism is responsible for the
high energy emission of the source over the whole observing period. 

\noindent Moreover, it is worth noting that the variability amplitude observed
at the end of March 2009 is similar in optical (a factor of $\sim$3) and
$\gamma$ rays (a factor of $\sim$3--4). If the $\gamma$ rays were produced by
inverse Compton scattering of photons reprocessed by the BLR, as suggested by
Abdo et al.~(2010a), the $\gamma$-ray emission would be proportional to the
number of the emitting electrons, like the synchrotron flux. Therefore, the
nearly linear correlation observed between optical and $\gamma$ rays may be
due to the variation of the electron number in the jet. In this context, the
lag of 1 day of the optical peak with respect to the $\gamma$-ray one could be
only an upper limit due to the lack of data before the observed optical peak.  

Compared to the optical and $\gamma$-ray activity, the X-ray flux is not very variable and does not seem to be correlated with the high $\gamma$-ray
activity (see Fig.~7, {\it panel c}). The lack of evident X-ray/$\gamma$-ray
correlation is not so surprising. In fact, Marscher et al.~(2010) already
reported that the long-term X-ray light curve of PKS 1510$-$089 collected
during the 2006--2009 period by {\it Rossi X-ray Timing Explorer} is
significantly correlated with the 14.5 GHz one rather than with the
$\gamma$-ray emission. This evidence could be justified by the fact that the
X-ray photons are likely to originate mostly in the low-energy part of the EC
emission. Therefore the X-ray spectral index seems to reflect the change in
the low-energy tail of the electron distribution, whereas a change in the
high-energy part of the electron distribution should be the driver of the
variation of the $\gamma$-ray flux, as already pointed out by Abdo et
al.~(2010a). Moreover, as already discussed in Kataoka et al.~(2008) and
D'Ammando et al.~(2009a), the spectral evolution of PKS 1510$-$089 could be
justified by the contamination at low energies of a second emission component
with respect to the EC emission. In this case the X-ray spectral shape of the
EC component should remain almost constant, regardless of the activity level,
but its flux should vary. Therefore the amount of contamination of the second
component should change, being more important when the source is fainter and
almost hidden by the EC component when the source is brighter. Possible
origins proposed for this second component are the soft X-ray excess, the bulk
Comptonization or a more significant contribution of the SSC component in any
activity states. A possible hint of increase of activity in X-rays is detected
by XRT on 28 March and this could be an indication of an enhancement of the
contribution of the synchrotron self Compton (SSC) emission, strictly related
to the optical/UV outburst observed on 27 March. The strength of the SSC with
respect to EC emission depends on the ratio between the synchrotron and the
external radiation energy density, as measured in the comoving frame, $U'_{\rm
  syn}/U'_{\rm ext}$. Taking into account that, under the assumption that the
dissipation region is within the BLR, $U'_{\rm syn}$ depends on the injected
power, the size of the emission and the magnetic field, a larger magnetic
field can lead to a larger SSC contribution together with an increase and a shift of the synchrotron peak. By contrast, an increase of the SSC
emission should lead to a not negligible contribution of this component, which
usually shows a photon index of $\Gamma$ $\sim$ 2 in X-rays for the FSRQs, and
therefore a softening of the total X-ray spectrum. Instead a hard X-ray
spectrum was observed by XRT ($\Gamma_{x}$ = 1.31 $\pm$ 0.14 on 28 March),
confirming that also during a contemporaneous synchrotron flare the SSC
contribution is not significant in the X-ray spectrum of PKS 1510$-$089, as
already observed by {\it Suzaku} during a low state of the source in 2009 (Abdo et al.~2010c).    

An alternative scenario is that the electrons radiating at 14.5 GHz produce
X-rays via IC scattering of the IR photons from a dust torus. This
interpretation favours the ``far dissipation'' scenario recently proposed
again for the emission of FSRQs by Sikora et.~(2008) and Marscher et
al.~(2010), in which the high energy peak is mainly due to the EC scattering
of the IR seed photons produced by the torus. In particular, Marscher et
al.~(2010) identified the dissipation zone with a moving superluminal knot
before and after passing the core, which is interpreted as the first
recollimation shock visible at millimeter wavelengths in the jet.

It is crucial to discriminate between these two scenarios in order to
understand where the plasma blob dissipates in the jet. In fact, the energy
density of the different sources of seed photons varies with the distance from
the central black hole, therefore the dominant population of target photons
depends on the site of the dissipation region that produces the flaring episode. The determination of the location of the emitting region in the jet of
blazars is an open issue, in particular for FSRQs. However, the rapid
variability observed in the $\gamma$-ray light curve of PKS 1510$-$089
collected by AGILE in March 2009 (even more remarkable in the {\it Fermi}-LAT
light curves on timescales of hours, Abdo et al.~(2010a), Tavecchio et
al.~2010) puts severe constraints on the size of the emitting region, leading
to a very compact region difficult to be explained if the $\gamma$-ray
emission is produced far away from the central black hole. This seems to
disfavour the ``far dissipation'' scenario. On the other hand, the recent
detection of this source at TeV energies by H.E.S.S. (Wagner and Behera 2010)
suggests possible complex scenarios. We can envisage that more than one
emission region is simultaneously active and located at different distance
from the central black hole in a spine-sheath (Tavecchio and Ghisellini 2008) or jet-in-jet structure (Giannios et al.~2009). 

It is worth noting that the outburst detected by {\it Swift}/BAT on 8--9 March
2009 occurred just at the beginning of the $\gamma$-ray activity observed by
AGILE. On the other hand, no significant activity is detected by BAT simultaneously with the $\gamma$-ray flares detected by AGILE.

Finally, as shown in Figure~\ref{radio}, no clear sign of strong activity in
  radio bands is observed during March 2009, although the radio coverage is
  sparse. However, the SMA data show an increase of about a factor of two
  between February and March 2009 at millimeter wavelengths, similarly to the
  optical behaviour even if with a minor variation amplitude. Moreover, the
  radio data acquired at Mets\"ahovi at 37 GHz on a longer timescale show that
  the increase of activity marginally observed also in Fig.~\ref{radio}, which
  starts at the beginning of March 2009 (JD $\sim$ 2454910), has continued
  gradually in April--May 2009, reaching a peak flux density of 4.02 Jy on 15
  May 2009 (JD 2454969.4; see Abdo et al.~2010a). The increase of flux density
  observed at 37 and 230 GHz confirms that the mechanism producing the
  $\gamma$-ray flaring events also interested the mm/cm emitting region with
  difference in time likely related to opacity effects, as already observed in
  March--April 2008 during the period soon after the previous $\gamma$-ray
  flare of PKS 1510$-$089 detected by AGILE (D'Ammando et al.~2009a). 

In addition, the VLBA data show a minimum of the flux density of the core in
March 2009 and since the rotation of the polarization angle of the core was
very mild in March relative to next 50 days, it is unlikely that the March
$\gamma$-ray flaring episodes are related to the the blob associated with
later flares as in the Marscher et al.~(2010) model. 

\section{Summary and concluding remarks}

We reported on the detection by AGILE-GRID of $\gamma$-ray activity from the
FSRQ PKS 1510$-$089 during March 2009. AGILE data showed an amazing
$\gamma$-ray activity, with several flaring episodes and a flux that reached
on daily integration $\sim$ 700 $\times$ 10$^{-8}$ photons cm$^{-2}$ s$^{-1}$,
the highest flux detected from this source until now and one of the highest detected from a blazar. 
No significant spectral changes are observed between the different
$\gamma$-ray flares in March 2009, indicating that the same mechanism is
dominant for the high energy emission of the source over all the period. 

This $\gamma$-ray activity triggered simultaneous multiwavelength
observations, which provided us a wide dataset for studying the correlation
between the emission properties from radio to $\gamma$ rays. An increasing
activity in near-IR and optical was observed, with a strong flux enhancement
after 24 March and a flaring episode on 26-27 March 2009 observed by GASP-WEBT
and REM, almost simultaneous to the brightest $\gamma$-ray flare. By contrast,
no clear signs of simultaneous high activity were detected from near-IR to UV
during the other $\gamma$-ray flares observed by AGILE. 

The broad band spectrum from radio-to-UV collected on 18 March confirmed the evidence of thermal signatures, the little and big blue bumps, in the
optical/UV spectrum of PKS 1510$-$089 also during high $\gamma$-ray states,
and not only during low activity states as observed in other blazars. The
radio-to-UV spectrum on 25-26 March showed a flat spectrum in the optical/UV
energy band, suggesting an important contribution of the synchrotron emission
in this part of the spectrum during the brightest $\gamma$-ray flare and
therefore a significant shift of the synchrotron peak, in agreement with the
rapid variation detected by UVOT and the drastic change of polarization angle observed in VLBA data simultaneously with the optical flare. The optical/UV
spectra presented here clearly show the different contribution of the thermal
and non-thermal components in the various phases of the $\gamma$-ray activity
of the source, with consequent theoretical implications on the modeling of the
broad band spectrum to investigate. It will be important to continue the
monitoring of the source from IR to UV to know if the shift of the synchrotron peak observed in March 2009 is a rare event or not.      

The {\it Swift}/XRT observations show no clear correlation between the X-ray and $\gamma$-ray emission. This could be due to the fact
that X-rays and $\gamma$-rays are originated by EC emission of different parts
of the electron distribution with different variability. On the other hand, a
hint of harder-when-brighter behaviour was detected from the XRT observations
suggesting the presence of a second component in the soft X-ray part of the
spectrum that could be associated with the soft X-ray excess rather than to
the variation of the SSC contribution in different activity states of the source. 

Moreover, the radio data show that an increase of the activity related to the $\gamma$-ray activity started in March and peaked in April-May 2009,
confirming that the mechanism producing the $\gamma$-ray flare interested also
the mm/cm emitting zone, with some delay likely due to opacity effects.

In conclusion, the $\gamma$-ray emission of PKS 1510$-$089 shows a complex
correlation with the other wavelengths, leading to different scenarios when
modeling its broad band spectrum at different epochs. The multifrequency
observations presented here give new clues, but also offer new questions on
the astrophysical mechanisms at work in this object. Further IR observations
can be important to test the ``far dissipation'' scenario in which the IR
photons are the dominant seed photons for the IC mechanism that produces the
$\gamma$-rays, and in this context the launch of Herschel (Pillbratt et
al.~2010) opened a new window on the infrared Universe and it will be very
important also for the study of blazars (Gonzalez-Nuevo et al.~2010). Finally,
the study of the correlation between the $\gamma$-ray activity and the
variability of optical polarization provided new information on the jet
structure, the location and the causes of the high-energy emission in PKS
1510$-$089. A further piece of the puzzle can be added with the launch of a X-ray polarimeter onboard any future X-ray missions (e.g.~the New Hard
X-ray Mission, Pareschi et al.~2010) giving us a final answer to the real
nature of the X-ray spectrum of this puzzling object.    

\begin{acknowledgements}

The AGILE Mission is funded by the Italian Space Agency (ASI) with scientific
participation by the Italian Institute of Astrophysics (INAF) and the Italian
Institute of Nuclear Physics (INFN). We thank the Swift Team for making these
observations possible, particularly the duty scientists and science planners. The Submillimeter Array is a joint project between the Smithsonian
Astrophysical Observatory and the Academia Sinica Institute of Astronomy and Astrophysics and is funded by the Smithsonian Institution and the Academia
Sinica. UMRAO is funded by a series of grants from the NSF and NASA and by the
University of Michigan. This research has made use of data from the MOJAVE
database that is maintained by the MOJAVE team (Lister et al.~2009, AJ, 137, 3718). Acquisition of the MAPCAT data is supported in part by MICIIN (Spain)
grants AYA2007-67627-C03-03 and AYA2010-14844, and by CEIC (Andaluc\'{i}a)
grant P09-FQM-4784. Abastumani Observatory team acknowledges financial support
by the Georgian National Science Foundation through grant GNSF/ST08/4-404. The
Liverpool Telescope is operated on the island of La Palma by Liverpool John
Moores University in the Spanish Observatorio del Roque de los Muchachos of
the Instituto de Astrofisica de Canarias, with funding from the UK Science and
Technology Facilities Council. The Mets\"ahovi team acknowledges the support
from the Academy of Finland to our observing projects (numbers 212656, 210338,
and others). St. Petersburg University team acknowledges support from Russian
RFBR foundation via grant 09-02-00092. This paper is partly based on
observations carried out at the German-Spanish Calar Alto Observatory, which
is jointly operated by the MPIA and the IAA-CSIC. Some of the authors
acknowledge financial support by the Italian Space Agency through contract ASI-INAF I/088/06/0 and I/009/10/0 for the Study of High-Energy Astrophysics.

\noindent {\it Facilities}: AGILE, {\it Swift}, WEBT, REM, UMRAO, and VLBA.

\end{acknowledgements}
%
%

\end{document}